\documentclass[pra,aps,showpacs,amsmath,amssymb,twocolumn,superscriptaddress]{revtex4}
\usepackage{color}             
\newcommand{\half}{\frac{1}{2}}         
\newcommand{\bla}{\color{black}}              \newtheorem{thm}{Theorem}

\begin{document}
\title{Complementarity between  signalling and local  indeterminacy in
  quantum nonlocal correlations}

\author{S.  Aravinda}\email{aru@poornaprajna.org}

\affiliation{Poornaprajna    Institute    of   Scientific    Research,
  Sadashivnagar, Bangalore, India.}

\author{R.  Srikanth}\email{srik@poornaprajna.org}

\affiliation{Poornaprajna    Institute    of   Scientific    Research,
  Sadashivnagar, Bangalore, India.}

\affiliation{Raman   Research  Institute,   Sadashivnagar,  Bangalore,
  India.}

\begin{abstract}
The correlations  that violate the  CHSH inequality are known  to have
complementary  contributions from  signaling and  local indeterminacy.
This complementarity  is shown to represent a  strengthening of Bell's
theorem, and can be used to certify randomness in a device-independent
way,  assuming neither  the  validity of  quantum  mechanics nor  even
no-signaling.  We obtain general  nonlocal resources that can simulate
the statistics of the singlet state, encompassing existing results. We
prove a  conjecture due to Hall (2010)  and Kar et al.   (2011) on the
complementarity for  such resources.
\end{abstract}

\maketitle

\section{Introduction}

Quantum correlations  are nonlocal in that they  can violate Bell-type
inequalities \cite{Bel64,CHSH},  which a local-realistic  model cannot
violate.  A 1-bit signal \cite{TB03}  or a single use \cite{CGM+05} of
the  Popescu-Rohrlich  (PR)  box  \cite{PR94}  can  reproduce  singlet
statistics. It  was shown by Kar  et al.  \cite{KGB+11}  that a convex
combination of  the above two  resources should also  simulate singlet
statistics, indicating  a trade-off between signaling  ($S$) and local
indeterminacy  ($I$)  in the  resources  used  for simulating  singlet
statistics.  Complementary  contributions from $S$ and  $I$ to quantum
correlations  have also been  studied in  Ref.  \cite{Hal10}.   In the
present  work, we  derive  a quantitative  relationship between  these
quantities, and  use it prove  a conjecture due to  Hall \cite{Hal10},
that  $S +  2I  \geq 1$  for  resources required  to simulate  singlet
statistics, as  well as the  entropic version of the  conjecture, that
$H_S  + H_I  \geq  1$ \cite{KGB+11},  where  $H_S$ and  $H_I$ are  the
corresponding entropic versions.

\section{Signaling, indeterminacy and communication cost: 
     Definitions \label{sec:defn}}

Bell's  theorem (or its  variants) says  that a  bipartite correlation
$P(ab|xy)$  generated  by local-realistic  theories  must satisfy  the
Clauser-Horne-Shimony-Holt (CHSH) inequality:
\begin{equation}
\Lambda = E(0,0) + E(0,1) + E(1,0) - E(1,1) \le 2,
\label{eq:bi}
\end{equation}
with  $a, b,  x, y  \in \{0,1\}$.   Here $  E(x,y) \equiv  P(a=b|xy) -
P(a\ne  b|xy)$.    More  generally,   it  applies  to   any  bipartite
correlation where outcomes $a,b$ are assumed to be pre-determined, and
$x, y$ are freely chosen  \cite{Hal10} and uncorrelated with the other
party's output.

A correlation  $\textbf{P} \equiv  P(a,b|x,y)$ is non-signaling  if it
satisfies:
\begin{eqnarray}
\sum_b  P(a,b|x,y=0) =  \sum_b P(a,b|x,y=1)  \equiv  P(a|x), \nonumber
\\ \sum_a P(a,b|x=0,y) = \sum_a P(a,b|x=1,y) \equiv P(b|y),
\label{eq:nosig}
\end{eqnarray}
i.e.,   Alice  knows  nothing   of  Bob's   input,  and   vice  versa,
respectively.   The   amount  of  signal from Alice to Bob and
Bob to Alice, respectively, can   be  quantified  either
statisitically as $S$ or entropically as $H_S$, as follows:
\begin{eqnarray}
S_{A\rightarrow     B}     &=&    \sup_{x,x^\prime,y,b}|P(b|x,y)     -
P(b|x^\prime,y)|\nonumber\\        S_{B\rightarrow        A}       &=&
\sup_{x,y,y^\prime,a}|P(a|x,y) - P(a|x,y^\prime),
\label{eq:defsig}
\end{eqnarray}
where  $P(a|x,y)   =  \sum_b   P(a,b|x,y)$  and  $P(b|x,y)   =  \sum_a
P(a,b|x,y)$. The signal
\begin{equation}
S = \max\{S_{A\rightarrow B}, S_{B\rightarrow A}\}
\end{equation}
The entropic version of quantity of signal is
\begin{equation}
H_S = \max\{\sup_x I(A:Y), \sup_y I(B:X)\},
\label{eq:entsig}
\end{equation}
where  $I(A:Y)$ denotes  mutual information  and $A,  B, X,  Y$  are random
variables representing $a, b, x, y$.

The  \textit{communication cost}  $C$ of  $\textbf{P}$ is  the minimum
size of a  classical message that must be  exchanged between Alice and
Bob in  a classical protocol  to reproduce $\textbf{P}$.   In general,
this message  must contain both  the input and outcome  information of
the other  party \cite{PKP+10}.   However, assuming that  both parties
have unrestricted  access to  shared randomness, and  that measurement
settings are chosen freely, the outcome information may be taken to be
determined  by the pre-shared  randomness.  Thus  it suffices  for the
communication  cost to  be large  enough to  convey just  the settings
information.   For the  two-input two-outcome  correlations considered
here, this is just 1 bit.   For example, the PR box is a non-signaling
resource that satisfies the condition
\begin{equation}
a \oplus b =x \cdot y,
\label{eq:prbox}
\end{equation}
thereby violating the CHSH  inequality to its algebraic maximum, going
beyond the Tsirelson bound \cite{cir80}. It is described by the action
\begin{equation}
P(a,b|x,y) = \left\{
\begin{array}{l l}
  \frac{1}{2}  & \textrm{Eq.~}(\ref{eq:prbox})~\textrm{holds}  \\  0 &
  \quad \mbox{otherwise}.\\ \end{array} \right.
\label{eq:nlb}
\end{equation}
The indeterminacy of $\textbf{P}$ can be quantified statistically as
\begin{equation}
I \equiv \sup_{x,y} \min_{o} \{P(o|x,y), 1-P(o|x,y)\},
\label{eq:indi}
\end{equation}
where  $o$  is  the outcome  on  any  one  of  the party's  side.   If
$\textbf{P}$ is interpreted operationally, i.e., $\textbf{P}$ is taken
to be the  correlation generated by measurements on  a physical state,
then   it   represents   \textit{unpredictability}  \cite{CW12}.    If
$\textbf{P}$ is interpreted as a  simulating resource or as an element
of   an  underlying   hidden-variable  theory,   then   it  represents
indeterminacy \cite{Hal10}, a term  which we also use generically here
to  describe  a  formal  correlation  $\textbf{P}$.   The  information
theoretic equivalent of $I$ may be given by the measure
\begin{equation}
H_I \equiv \sup_{x,y} H(O|x,y),
\label{eq:indient}
\end{equation}
where $H(O|x,y) \equiv - \sum_{o} p_o\log_2(p_o)$. 

\section{Interplay of signaling and indeterminacy in nonlocal correlations
\label{sec:theorem}}
A   correlation  $\textbf{P}$   generated  by   two-input,  two-output
bi-partite measurements on a physical state, or which can be used as a
resource to reproduce such correlations, can be decomposed as a convex
combination  of  deterministic  correlations  or  `boxes'  (for  which
$P(a,b|x,y)  = 0$ or  1) that  are 1-bit  strategies, having  the form
$P(a,b|x,y)  =  \delta^{a}_{f(x,y)}\delta^b_{g(y)}$  or $P(a,b|x,y)  =
\delta^{a}_{f(x)}\delta^b_{g(x,y)}$ (with  $C=1$) or 0-bit strategies,
having   the  form  $P(a,b|x,y)   =  \delta^{a}_{f(x)}\delta^b_{g(y)}$
($C=0$) \cite{Pir03}.

We may  uniformly average some pairs  of the above  signaling boxes to
create non-signaling correlations.  For  example, a uniform average of
$P^1_{+}  \equiv  P(ab|xy)  =  \delta^a_0\delta^b_{xy}$  and  $P^1_{-}
\equiv P(ab|xy) = \delta^a_{1}\delta^b_{xy \oplus 1} $, results in the
PR   box   (\ref{eq:nlb}).    We   call  pairs   like   $P^1_\pm$   as
\textit{signaling   pairs},  with   $P^1_\pm$   the  \textit{signaling
  complements}   of  $P^1_\mp$.    By  averaging   signal  complements
non-uniformly,  we  obtain  resources  of intermediate  signaling.   A
complete  listing  of the  deterministic  signaling correlations  that
satisfy  the PR  box  condition (\ref{eq:prbox})  are  given in  Table
\ref{tab:1way}.   The no-signaling polytope  has 8  nonlocal vertices,
corresponding to the PR boxes, characterized by the three bits $\mu_1,
\mu_2, \mu_3$, which define the general  PR box relation $a \oplus b =
x\cdot y \oplus \mu_1x \oplus \mu_2y \oplus \mu_3$ \cite{BLM+05}.

\begin{widetext}
\begin{center}
\begin{table}
\begin{tabular}{c||c|c|c|c|c|c|c|c|c|c|c|c|c|c|c|c}
\hline  Input &  $S^1_+$  & $S^1_-$  &  $S^2_+$ &  $S^2_-$ &  $S^3_+$&
$S^3_-$ & $S^4_+$ & $S^4_-$ &  $S^5_+$ & $S^5_-$ & $S^6_+$ & $S^6_-$ &
$S^7_+$ & $S^7_-$ & $S^8_+$ & $S^8_-$ \\ \hline 00 & 00 & 11 & 00 & 11
& 00 & 11 & 00 & 11 & 00 & 11 & 00 & 11 & 00 & 11 & 00 & 11 \\ 01 & 00
& 11 & 00 & 11 & 00 & 11 & 11 &  00 & 11 & 00 & 00 & 11 & 11 & 00 & 11
& 00 \\ 10 & 00 & 11 & 00 & 11 & 11 & 00 & 00 & 11 & 00 & 11 & 11 & 00
& 11 & 00 & 11 & 00 \\ 11 & 01 & 10 & 10 & 01 & 10 & 01 & 01 & 10 & 10
& 01 & 01 & 10 & 01 & 10 & 10 & 01 \\ \hline
\end{tabular}
\caption{Table of deterministic  correlations in the \textit{scope} of
  the PR  box with $\mu_j=0$.  The first  eight columns, corresponding
  to deterministic  1-bit strategies  (i.e., having $C=1$),  are 1-way
  signaling, and the remaining are  2-way signaling.  The usual PR box
  (with $\mu_j=0$) is an equal weight convex combination of $S^1_\pm$,
  while the signaling resource $S^p$ considered in Ref.  \cite{KGB+11}
  corresponds to $\textbf{P} = pS^1_+ + (1-p)S^1_-$.}
\label{tab:1way}
\end{table}
\end{center}
\end{widetext}

A general decomposition of  a two-input two-output correlation, with a
possible signal either from Alice to Bob or vice versa, is given by:
\begin{equation}
\textbf{P}  = CS_1 + (1-C)S_0,
\label{eq:gencorr}
\end{equation}
where $C$ is communication cost,  $S_1$ is the nonlocal part (given as
a mixture of 1-bit strategies) and $S_0$ is the local part (given by a
mixture of 0-bit strategies).
\begin{thm}
For correlation \textbf{P} in Eq. (\ref{eq:gencorr})
\begin{equation}
S + 2I \ge C.
\label{eq:main}
\end{equation}
\label{thm:this}
\end{thm}
\textbf{Proof sketch.} We first consider simulating \textbf{P} that is
simulable using strategies in the scope (PR box) $\mu_j=0$.  We do not
require  individual   signal  pairs   to  be  balanced.    From  Table
\ref{tab:1way},  it is  seen that  Bob  receives a  signal from  Alice
setting $y=0$, when the strategies are $S^3_\pm, S^6_\pm, S^7_\pm$ and
$S^8_\pm$. The  probabilities of these strategies  thus determines the
signal  $S^{A  \rightarrow  B}_{y=0}$  in  the  resources.   Thus,  by
imbalancing this and  the other 4 signal complements  and denoting the
signal in each case by $s_k$, we have
\begin{widetext}
\begin{eqnarray}
S^{A\rightarrow   B}_{y=0}  &=&   \left(p^3_+  +   p^6_+  +   p^7_+  +
p^8_+\right) - \left(p^3_-  + p^6_- + p^7_- +  p^8_-\right) \equiv s_1
\nonumber \\ S^{A\rightarrow B}_{y=1}  &=& \left(p^1_+ + p^5_- + p^6_+
+ p^8_-\right) - \left(p^1_- + p^5_+ + p^6_- + p^8_+\right) \equiv s_2
\nonumber \\ S^{B\rightarrow A}_{x=0}  &=& \left(p^4_+ + p^5_+ + p^7_+
+ p^8_+\right) - \left(p^4_- + p^5_- + p^7_- + p^8_-\right) \equiv s_3
\nonumber \\ S^{B\rightarrow A}_{x=1}  &=& \left(p^2_+ + p^5_+ + p^6_-
+ p^7_-\right)  - \left(p^2_-  + p^5_- +  p^6_+ +  p^7_+\right) \equiv
s_4.
\label{eq:4sig+}
\end{eqnarray}
\end{widetext}
Therefore, $\sum_{j=1}^8 (p^j_{+} - p^j_{-}) = s_1 + s_2 + s_3 + s_4$.
Since \textbf{P}  in (\ref{eq:gencorr}) has  non-vanishing probability
only  in 1-way  strategies, and  thus its  communication cost  \bla is
$C=\sum_{j=1}^4 \left(p^j_{+} + p^j_{-}\right)$, it follows that
\begin{eqnarray}
 \sum_{j=1}^4 
p^j_{+} & = & \frac{C + s_1 + s_2 + s_3 + s_4}{2} \nonumber\\ 
 \sum_{j=1}^4 p^j_{-} & = & \frac{C - s_1 - s_2 - s_3 - s_4}{2}
\end{eqnarray}
From Table \ref{tab:1way}, we have $P(00|00) \ge \sum_{j=1}^4 p^j_{+}$
and $P(11|00) \ge \sum_{j=1}^4 p^j_{-}$, so that
\begin{eqnarray}
P(00|00) & \ge & \frac{C + s_1 + s_2 + s_3 + s_4}{2} \nonumber\\ 
P(11|00) & \ge & \frac{C - s_1 - s_2 - s_3 - s_4}{2}.
\label{eq:singsimu}
\end{eqnarray}
The inequalities above follow from  the fact that $P(00|00)$ etc.  may
have contributions also from the local strategies. (If $C=1$, we would
have equalities here.)   By the same method we  have all the remaining
conditional probabilities
\begin{eqnarray}
P(00|01) &\ge& \frac{C  + (s_1 + s_2 - s_3 + s_4)}{2} \nonumber \\
P(11|01) &\ge& \frac{C-(s_1 + s_2 - s_3  + s_4)}{2} \nonumber \\
P(00|10) &\ge& \frac{C
    + (-s_1 + s_2 + s_3  + s_4)}{2} \nonumber \\
P(11|10) &\ge& \frac{C- (-s_1 +
    s_2 + s_3 + s_4)}{2} \nonumber \\
P(01|11) &\ge& \frac{C + (-s_1 + s_2
    + s_3 - s_4)}{2} \nonumber \\
P(10|11) &\ge& \frac{C - (-s_1  + s_2 + s_3 -
    s_4)}{2}.
\label{eq:singsimu0}
\end{eqnarray}
Let us consider  case [A] $s_1 \leq  s_2 \leq s_3 \leq s_4  $ and [A1:]
$s_1 + s_4 > s_2 + s_3$. From definition (\ref{eq:indi}), we have
\begin{equation}
I \ge \frac{C + (-s_1 + s_2 + s_3 - s_4)}{2},
\end{equation}
and Ineq.  (\ref{eq:main}) using  assumption [A1].  If we consider the
case [A2:] $s_1 + s_4 < s_2 + s_3$, then
\begin{equation}
I \ge \frac{C - (-s_1 + s_2 + s_3 - s_4)}{2}
\end{equation}
from which,  once again, Eq. (\ref{eq:main})  follows, using condition
[A2].    Repeating   the  above   exercise   for   all  other   cases,
Eq. (\ref{eq:main}) is  seen to hold in a  similar fashion. Since each
scope  (i.e., PR  box) can  be converted  to another  using reversible
local operations \cite{BLM+05}, the  result holds true for any mixture
of the scopes.  \hfill $\blacksquare$

For   an  arbitrary  nonlocal   correlation  \textbf{P},   our  result
(\ref{eq:main}) implies
\begin{equation}
S + 2I > 0.
\label{eq:BI}
\end{equation}
Eq.   (\ref{eq:BI}) can be  interpreted as  an operational  version of
Bell's    inequality,    derived     under    the    assumptions    of
\textit{signal-locality}  ($S=0$) and  \textit{predictability} ($I=0$)
\cite{CW12}. Our result Eq.  (\ref{eq:main}) is then seen to represent
a strengthening of Bell's theorem, Eq.  (\ref{eq:BI}).

\section{Certified randomness \label{sec:cert}}

Randomness, while  very important in  modern science and  industry for
simulations, is nevertheless an elusive concept \cite{Cha82}.  Given a
purported  source of  randomness,  it is  difficult  to ascertain  its
random  nature  without  characterizing  the  detailed  structure  and
mechanism behind it. Randomness certified by Bell's theorem provides a
way out of this  difficulty \cite{PAM+10,DTA14}.  If Bell's inequality
is violated by the observed correlation \textbf{P} between two distant
parties,  Alice and Bob,  whose measurements  are spacelike-separated,
then  as signaling is  fundamentally disallowed,  Eq.  (\ref{eq:main})
implies  that  there  is  an  irreducible  randomness  in  \textbf{P},
irrespective of a detailed characterization of the devices used.  Thus
a bound on  randomness obtained by a Bell  test is device-independent.
Our above results can be used to generalize this idea in two ways: one
is  that  quantum  mechanics  is  not assumed,  and,  further  nor  is
no-signaling.

It   is  known   that  $C   \ge  \frac{\Lambda(\textbf{P})}{2}   -  1$
\cite{Pir03}.  Substituting this in Eq. (\ref{eq:main}), we find:
\begin{equation}
I \ge \frac{\Lambda(\textbf{P})}{4} - \frac{1 + S}{2},
\label{eq:cert}
\end{equation}
as the amount  of randomness certified by a Bell  test in the presence
of  signaling. Intuitively,  the greater  the signal,  the  larger the
classical  explanation for a  Bell's inequality  violation \cite{AS2},
and hence lower the certifiable randomness.

Rewriting Eq. (\ref{eq:cert}), we obtain a version of the
relaxed Bell's inequality
\begin{equation}
\Lambda(\textbf{P}) - 2 \le 2S + 4I,
\label{eq:relax}
\end{equation}
where  the amount  of  CHSH  inequality violation  (in  the l.h.s)  is
bounded by the signaling and  indeterminacy in the correlation (cf.  a
similar result in Ref. \cite{Hal10}).

\section{Compelementarity between signaling and indeterminacy
in simulating  singlet statistics \label{sec:signindet}}  

If the correlation $\textbf{P}$ is  used as a resource to simulate the
correlations   in  a  physical   theory,  then   Eq.   (\ref{eq:main})
represents  the complementarity  for the  simulating  resources.  Now,
modelled as  a mixture of local and  nonlocal strategies, correlations
representing a singlet have no local contribution \cite{BKP06}.  Thus,
consider as a resource the general signaling, nonlocal box obtained by
the convex combination of the 1-bit strategies of Table \ref{tab:1way}
\begin{equation}
\textbf{P} =  \sum_{j=1}^4 (p^j_+S^j_+ + p^j_-S^j_-),
\label{eq:genP}
\end{equation}
where $\sum_{j=1}^4(p^j_+ + p^j_-)=1$. The protocol of Toner and Bacon
\cite{TB03}  corresponds  to the  case  of  setting  all $p^j_\pm$  in
Eq.  (\ref{eq:genP}) to  0 except  one (say,  $p^1_+=1$).  The  PR box
simulation of  Cerf et al.   \cite{CGM+05} corresponds to the  case of
setting  all  $p^j_\pm$ in  Eq.   (\ref{eq:genP})  to  0 except  those
belonging to one signaling pair, which are both equally weighted (say,
$p^1_+=p^1_- = \half$).  The  more general simulation presented by Kar
et al.  \cite{KGB+11} corresponds to the case of setting all $p^j_\pm$
in Eq. (\ref{eq:genP}) to 0  except those belonging to one pair, which
now are not required to be  equally weighted (say, $p^1_+ + p^1_- = 1$
and $p^1_+ \ne p^1_-$).  In our notation, all these nonlocal resources
belong   to   the   same   signaling   pair.    Our   result   follows
straightforwardly from  the observation that  the simulation protocols
of  Refs.    \cite{CGM+05,  KGB+11}  work  even   when  \textbf{P}  is
generalized as in Eq.  (\ref{eq:genP}) with unrestricted signal domain
in  the same  PR scope,  essentially  because each  of the  underlying
deterministic   strategies    considered   satisfies   the   condition
(\ref{eq:prbox}).   A  general resource  of  the type  (\ref{eq:genP})
drawn  from any  other,  fixed  scope (a  different  triple of  values
$\mu_j$)  would also  do, since  the different  PR boxes  are mutually
transformable through reversible local relabelling.

For completeness, we give the full protocol that simulates the singlet
state correlation using  resource \textbf{P} and pre-shared randomness
$\hat{\theta}_1$  and $\hat{\theta}_2$,  which  are independently  and
uniformly distributed directional vectors. Alice (Bob) is given vector
$\hat{x}$   ($\hat{y})$  and   outputs   binary  number   $\textbf{x}$
($\textbf{y}$) taking  value 0 or  1. To simulate  singlet statistics,
they must satisfy:
\begin{equation}
 \overline{\textbf{x} \oplus \textbf{y}  |  \hat{x}, \hat{y}  }  = \frac{1  +
   \hat{x} \cdot \hat{y}}{2},
\label{eq:singlet}
\end{equation}
where  the overline  indicates the  expectation value.   To  this end,
Alice computes  $x = \mbox{sgn}(\hat{x} \cdot  \hat{\theta}_1 ) \oplus
\mbox{sgn}(\hat{x} \cdot  \hat{\theta}_2)$, which she  inputs into the
resource $\textbf{P}$.  Here sgn($z$)  = 0 (1)  if $z <  0$ ($z\ge0$).
Using output $a$ from the resource, Alice obtains:
\begin{equation}
\textbf{x} = a\oplus \mbox{sgn}(\hat{x}\cdot\hat{\theta}_1).
 \end{equation}
Bob computes the quantity $ y = \mbox{sgn}(\hat{y}\cdot\hat{\theta}_+)
\oplus          \mbox{sgn}(\hat{y}\cdot\hat{\theta}_-)$,         where
$\hat{\theta}_{\pm}=\hat{\theta}_1\pm   \hat{\theta}_2$,  which  input
into $\textbf{P}$, produces output $b$. Bob uses this to compute:
\begin{equation}
\textbf{y} = b\oplus \mbox{sgn}(\hat{y}\cdot\hat{\theta}_+)\oplus1 .
 \end{equation}
This yields
\begin{eqnarray}
\textbf{x}\oplus   \textbf{y}    &   =    &   a   \oplus    b   \oplus
\mbox{sgn}(\hat{x}\cdot\hat{\theta}_1)                          \oplus
\mbox{sgn}(\hat{y}\cdot\hat{\theta}_+)\oplus1  \nonumber \\
&=&
\sum_j        (P^j_{+}        +        P^j_{-})       xy        \oplus
\mbox{sgn}(\hat{x}\cdot\hat{\theta}_1)                          \oplus
\mbox{sgn}(\hat{y}\cdot\hat{\theta}_+)\oplus1  \nonumber\\  &=&  xy
\oplus          \mbox{sgn}(\hat{x}\cdot\hat{\theta}_1)         \oplus
\mbox{sgn}(\hat{y}\cdot\hat{\theta}_+)\oplus1,
 \end{eqnarray} 
from  which  Eq.  (\ref{eq:singlet})   follows  using  the  method  of
Ref. \cite{CGM+05}.

Now, 1  bit is sufficient to  simulate the singlet,  since the general
resource (\ref{eq:genP}) has a communication cost of 1 bit.  That this
is  also necessary  \cite{BKP06} follows  from the  optimality  of the
Toner-Bacon   protocol.    Accordingly,    we   set   $C=1$   in   Eq.
(\ref{eq:main}), obtaining the complementary relation
\begin{equation}
S + 2I \ge 1
\label{eq:main+}
\end{equation}
for signal and indeterminacy  contributions in correlations in singlet
statistics.   This  was  conjectured  by  Hall  \cite{Hal10}.   If  we
consider a non-signaling  model of quantum mechanics, we  set $S=0$ in
Eq.  (\ref{eq:main+}), so that $I = \half$.  Thus, 1 bit of randomness
can be certified using singlets (cf. \cite{PAM+10}).

To obtain  the entropic  version of the  above, we note  that entropic
indeterminacy is, using Eq.  (\ref{eq:indient}), just
\begin{equation}
H_I  \equiv -I\log_2(I)  - (1-I)\log_2(1-I)
\label{eq:detent}
\end{equation}
For a model with  signal $S$ from Alice to Bob, there  is a setting of
Bob such  that the  probability of an  outcome, $p$, shifts  to $p+S$,
when Alice  toggles her input. Thus,  the entropic signal  is given by
$H_S =  H\left(p + \frac{S}{2}\right) -  \frac{1}{2}H(p) - \frac{1}{2}
H(p+S)$, from  which it follows, by optimizing  over $p$ \cite{Hal10},
that
\begin{equation}
H_S \ge 1 - H\left(\frac{1-S}{2}\right).
\label{eq:entsign}
\end{equation}
From Eqs. (\ref{eq:detent}) and (\ref{eq:entsign}), we have
\begin{equation}
H_S + H_I \ge 1,
\end{equation}
conjectured by Hall \cite{Hal10} and Kar et al. \cite{KGB+11}.

\section{Discussions \label{sec:conclu}}

The  complementarity   of  contributions  from   signaling  and  local
indeterminacy  to  nonlocal correlations  was  derived,  and shown  to
represent a strengthening of Bell's theorem. Our result, which applies
to arbitrary  degrees of violation  of Bell's inequality, was  used to
verify  a  conjecture  about  the  complementarity  in  the  resources
required to simulate singlet statistics.  Finally we obtain a bound on
the  randomness that  can  be  certified by  nonlocality  even in  the
presence of signaling.

The complementarity  (\ref{eq:main+}) unifies  a number of  results on
the  simulation  of  singlet  statistics.   Leggett  \cite{Leg03}  and
Gr\"oblacher et al.   \cite{GPK+07} proposed non-signaling models with
local bias,  which were shown  to be incapable of  reproducing singlet
statistics.   Local   bias  is   equivalent  in  our   terminalogy  to
$I<\frac{1}{2}$,  and since $S=0$  here, such  models fail  to satisfy
Ineq. (\ref{eq:main+}).  Thus complementarity explains why such models
fail to simulate singlet  statistics.  It also provides an alternative
proof of the result obtained  by Branciard et al.  \cite{BBG+08}, that
any  non-signaling  model of  singlet  statistics  must have  unbiased
marginals ($I=\frac{1}{2}$).
 
\acknowledgments

SA acknowledges  support through the INSPIRE  fellowship [IF120025] by
the Department of  Science and Technology, Govt. of  India and Manipal
University  graduate programe.  RS acknowledges  support from  the DST
project SR/S2/LOP-02/2012.

\bibliography{quantarv}
 
\end{document}